\def\aap{\mbox{\bf{Astron.\ Astrophys.}}}
\def\apj{\mbox{\bf{Astrophys.\ J.}}}
\def\apjl{\mbox{\bf{Astrophys.\ J.\ L.}}}
\def\aj{\mbox{\bf{Astron.\ J.}}}
\def\aaps{\mbox{\bf{Astron.\ Astrophys.\ Suppl.}}}

\def\mnras{\mbox{\bf{Mon.\ Not.\ Roy.\ Astron.\ Soc.}}}

\def\apjs{\mbox{\bf{Astrophys.\ J.\ Suppl.}}}

\newcommand{\msun}{\mbox{$M_{\odot}$}}

\newcommand{\mmax}{\mbox{$M_{\rm max}$}}

\newcommand{\sex}{SExtractor~}
\newcommand{\dr}{\mbox{${\rm d}$}}
\newcommand{\B}{\mbox{{\it F435W}}}
\newcommand{\V}{\mbox{{\it F555W}}}
\newcommand{\I}{\mbox{{\it F814W}}}
\newcommand{\Ha}{\mbox{{\it F658N}}}
\newcommand{\reff}{\mbox{{$R_{\rm eff}$}}}
\newcommand{\redchisq}{\mbox{$\chi_{\nu}^2$}}

\documentclass{aa}
\usepackage{graphicx}
\usepackage{natbib}
\bibpunct{(}{)}{;}{a}{}{,} 

\begin{document}
   \title{Observational evidence for a truncation of the star cluster initial mass function at the high mass end}

     \titlerunning{Truncation of the star cluster mass function}

   \author{M. Gieles \inst{1} \and S.S. Larsen \inst{2} \and R.A. Scheepmaker \inst{1} \and N. Bastian \inst{3} \and M.R. Haas \inst{1} \and   H.J.G.L.M. Lamers \inst{1,4}   }

   \offprints{gieles@astro.uu.nl}

   \institute{$^1$Astronomical Institute, Utrecht University, 
              Princetonplein 5, NL-3584 CC Utrecht The Netherlands \\
              $^2$European Astronomical Institute, Karl-Schwarzchild-Strasse 2
              D-85748 Garching b. Munchen, Germany \\
	      $^3$Department of Physics and Astronomy, University College London,
	      Gower Street, London, WC1E 6BT\\
             $^4$SRON Laboratory for Space Research, Sorbonnelaan 2,
             NL-3584 CA Utrecht, The Netherlands\\
             }

   \date{Received  September 12, 2005; accepted December 8, 2005 }


   \abstract{We present the luminosity function (LF) of star clusters
   in M51 based on {\it HST/ACS} observations  taken as part of
   the Hubble Heritage project. The clusters are selected based on
   their size and with the resulting 5\,990 clusters we present one of
   the largest cluster samples of a single galaxy. We find that the
   LF can be approximated with a double power-law distribution with a
   break around $M_V = -8.9$. On the bright side the index of the
   power-law distribution is steeper ($\alpha = 2.75$) than on the
   faint-side ($\alpha = 1.93$), similar to what was found earlier for
   the ``Antennae'' galaxies. The location of the bend, however,
   occurs about 1.6 mag fainter in M51. We confront the observed
   LF with the model for the evolution of integrated properties of
   cluster populations of \citet{gieles05b}, which predicts that a
   truncated cluster initial mass function would result in a bend in,
   and a double power-law behaviour of, the integrated LF. The
   combination of the large field-of view and the high star cluster
   formation rate of M51 make it possible to detect such a bend in the
   LF.  Hence, we conclude that there exists a fundamental upper limit
   to the mass of star clusters in M51.  Assuming a power-law cluster
   initial mass function with exponentional cut-off of the form
   $N\,\dr M \propto M^{-\beta}\,\exp(-M/M_C)\,\dr M$, we find that
   $M_C = 10^5\,\msun$. A direct comparison with the LF of the
   ``Antennae'' suggests that there $M_C = 4\times10^5\,\msun$.  }

%
%
         
\maketitle

\section{Introduction}

There is a relation between the luminosity of the brightest star
cluster in a galaxy and the total number of clusters
(\citealt{2003dhst.symp..153W};
\citealt{2002AJ....124.1393L}), suggesting that sampling statistics is
determining the luminosity of the most luminous cluster. Since the
luminosity of clusters is heavily dependent on the age, a
straightforward translation from most luminous to most massive is not
possible.  Recently, \citet{2003AJ....126.1836H} showed that the
maximum cluster mass increases with log(age/yr) in the LMC and SMC,
which can be interpreted as a size-of-sample effect.  Also
\citet{2004MNRAS.350.1503W} suggest that the maximum cluster mass in a
galaxy depends on the star formation rate in the galaxy, hence the
total number of clusters. This suggests that it would be {\it
physically} possible to form a super massive cluster such as W3 in
NGC~7252 with a mass of $8\times10^7\,\msun$
\citep{2004A&A...416..467M} in our Milky Way, but the chance is just
very small. This issue is still heavily under debate and is subject of
this study.

The cluster luminosity function (LF) is a powerful tool for the study of
star cluster populations. In a wide variety of environments the LF can
often be well approximated by a power-law distribution: $N\,\dr
L~\propto~L^{-\alpha}\dr L$, where the index $\alpha$ is between 1.8 and
2.4 (e.g. \citealt{2002AJ....124.1393L};
\citealt{2003MNRAS.343.1285D}). The shape of the LF is related to, but
not necessarily identical to the cluster initial mass function
(CIMF). It is important to note that it is hard to relate the observed
LF directly to the underlying CIMF, since the LF contains clusters of
different ages. A star cluster fades about 5 magnitudes in 1 Gyr in
the $V$-band, which makes it hard to estimate the mass without knowing
the age. 

The LF of clusters in the ``Antennae'' galaxies
\citep{1999AJ....118.1551W}, however, is much better approximated by a
{\it double} power-law distribution. The bright side ($M_V\,\la\,-10$)
has a steeper slope ($\sim -2.7$) than the faint side ($\sim -2$). The
latter is close to the value found for other galaxies. This double
power-law nature with a {\it bend}, was interpreted by the authors as
a turn-over in the {\it mass} function. 

In \citet{gieles05b} we compared a cluster population model with
various observed luminosity functions from the literature. We
investigated various possible ways of detecting a truncated cluster
initial mass function and the possible biases caused by extinction,
disruption, variations in the cluster formation rate, etc. We concluded
that a truncated CIMF will be observed as a bend in the
integrated cluster luminosity function. We showed that tentative hints
for a truncation are present in NGC~6946 (from
\citealt{2002AJ....124.1393L}) and M51 (from
\citealt{2005A&A...431..905B}) and are clearly not present in the SMC
and the LMC (from \citealt{2003AJ....126.1836H}). 

 In this work we present a greatly improved LF of clusters in M51,
based on recently released deep {\it HST} observations with the {\it
Advanced Camera of Surveys (ACS)} covering the entire disk of M51. The
great resolution of the {\it ACS} camera is exploited by selecting
clusters based on their size. With this we are able to accurately
select clusters, even when they are as faint as individual bright high
mass stars. The improved resolution and larger field-of-view make it
possible to confirm the suggestion of \citet{gieles05b} that the LF of
M51 is of a double power-law nature.

We show that the bend in the LF is not necessarily related to a
corresponding turn-over in the MF, but results naturally if the CIMF
is a power-law distribution truncated at the high-mass end.

In \S~\ref{sec:data} we describe the data, source selection and
photometry. In \S~\ref{sec:lf} we present the LF in the three
available {\it ACS} filters of all extended objects in M51. A comparison
with the model is done in \S~\ref{sec:model} and a discussion and the
conclusions are presented in
\S~\ref{sec:conclusions}.

\section{Source selection and photometry}
\label{sec:data}

\subsection{Data}
We used the new {\it HST/ACS} (Hubble Heritage) data of M51 (NGC~5194) and its companion
NGC~5195 in \B\ (2720 sec.), \V\ (1360 sec.), \Ha\ (2720 sec.)  and
\I\ (1360 sec.). Six pointings, corresponding to
$430\times610^{\prime\prime}$ (=$17.5\times24.8\,$kpc), cover the
entire disk of M51 plus the region with NGC~5195. For details on
reduction we refer to
\citet{2005AAS...206.1307M} and the M51 mosaic website
(http://archive.stsci.edu/prepds/m51/).


\subsection{Source selection}
Source selection was done with the \sex package
(\citealt{1996A&AS..117..393B}; version 2.3.2). A background map was
made, by computing a mean and standard deviation of every section of
the image with a user defined size. Deviating pixels were iteratively
discarded until every pixel value of the background was within $\pm
3\sigma$ of the mean value.  Every area of at least three adjacent
pixels that exceeded the background by at least $5\sigma$ was called a
source. For the details of the background estimation and source
selection we refer to \citet{1996A&AS..117..393B}. The \B, \V\, and
\I\, coordinates of the sources were cross-correlated and only sources
within a two pixel uncertainty were kept. This resulted in a list of
75436 sources.

\subsection{Photometry}
Aperture photometry was performed on the source list, using the {\it
IRAF/DAOphot} package using a 5 pixel aperture radius and a background
annulus with an inner radius of 10 pixels and a width of 3
pixels. Aperture correction from a 5 pixel aperture to 10 pixels
(=0.5'') were measured on artificial sources, generated by the {\it
BAOlab} package (\citealt{1999A&AS..139..393L};
\citealt{2004A&A...416..537L}). A Moffat profile
\citep{1969A&A.....3..455M} with power-law index of $-1.5$ and an
effective radius ($\reff$),  which is the radius containing half the
 cluster light in projection, of 3 pc was convolved with the filter
dependent point spread function (PSF) of the {\it ACS} camera and the
aperture correction was measured. The PSF we used was observationally
determined from a crowded star field on a drizzled image of the
Galactic globular cluster 47 Tuc. For each filter a separate PSF was
determined. The resulting aperture corrections in \B, \V\, and \I\,
were $-0.16, -0.16$ and $-0.17$ mag, respectively. These values would
be 0.04 lower/higher for sources which are 1 pc bigger/smaller. The
aperture corrections between 0.5'' and infinity were taken from
Table~5 of
\citet{2005astro.ph..7614S}. Finally, a correction for
Galactic foreground extinction of $E(B-V) = 0.038$ was applied
according to Appendix B of \citet{1998ApJ...500..525S}. We did
completeness tests by adding the same artificial clusters as used for
determining the aperture correction to the image. A high background
part of the image was used and the resulting 90\% completeness limits
in \B, \V\ and \I\ were 23.3, 23.3 and 23.0 mag respectively.

\subsection{Radius measurements}
To distinguish between stars and clusters, we exploit the resolution
of the {\it ACS} camera (1 pixel = 0.05'') to measure the radii of all
sources detected in \B, \V\ and \I\ using the {\it ISHAPE} routine
within the {\it BAOlab} package. In summary, analytic profiles with
variable effective radii are convolved with the PSF, and are
then fitted to each source in the data. The best-fit
\reff\ was determined by minimizing the $\chi^2$. We choose a Moffat
profile  with a power-law index of $-1.5$. Comparing the
\redchisq\ of a point source (= PSF) fit to an extended profile fit,
 indicated whether a source is resolved. {\it ISHAPE} is able to pick
 up sources with a FWHM of 0.2 pixels, which corresponds to a \reff =
 0.5 pc at the distance of M51 (distance modulus = 29.62,
 \citealt{1997ApJ...479..231F}) The source detection and radius fits
 are used to define a source to be a cluster when
\vspace{-0.25cm}
\begin{enumerate}
\item the source is detected in \B, \V\ and \I,
\item the source magnitude is above the 90\% completeness limit,
\item the sources is extended, defined as $\reff > 0.5$ pc,
\item the $\redchisq$ of the extended profile fit is lower than that of the pure  PSF fit.
\end{enumerate}
\vspace{-0.25cm}
These criteria resulted in $5\,990$ clusters brighter than $V = 23.3$
($M_V = -6.32$). 

\section{The luminosity function}
\label{sec:lf}

\begin{figure*}[!t]
\begin{center}
    \includegraphics[height=4.6cm]{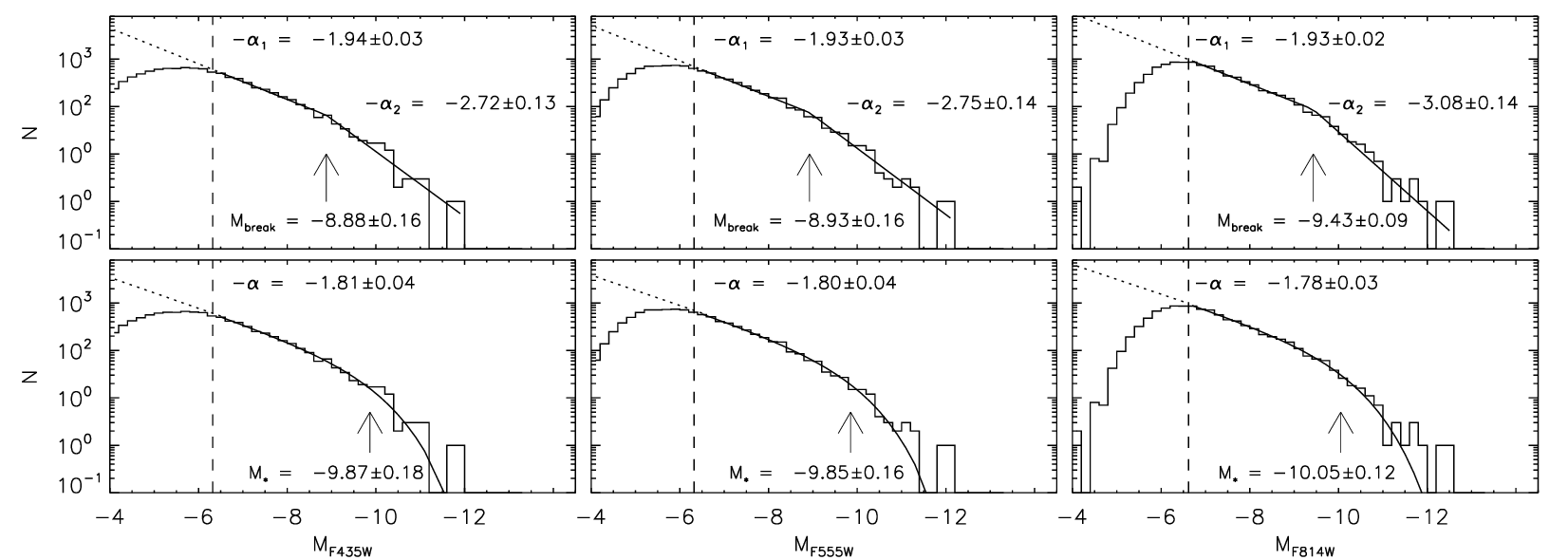}
    \includegraphics[height=4.6cm]{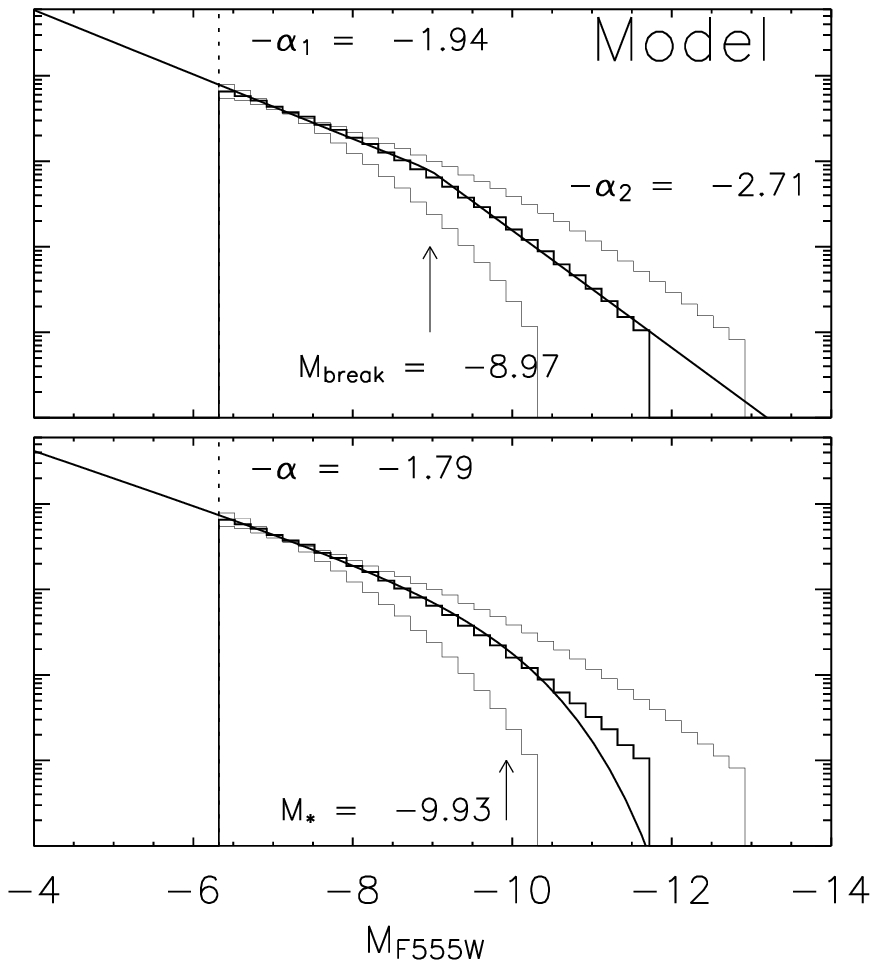}

    \caption{Luminosity function of all extended sources
    in M51 in \B, \V\ and \I\ ({\bf left three panels}) . The solid lines are the
    double power-law fits (top) and 
    Schechter fits (bottom). The typical bend is indicated as $M_*$ for the
    Schechter fits and $M_{\rm break}$ for the double power-laws. Also
    the slopes are indicated. The dotted line is the extrapolated
    distribution of the faint end. {\bf Right:} LF of a synthetic
    cluster population with a CIMF with exponent $-2$, a mass function
    truncation at $M_C = 10^5 \msun$ and the same detection
    limit as the data. The top panel shows a double power-law fit,
    where the slope bright-ward of the bend is steeper than the
    underlying mass function. The bottom panel shows a fit with a
    Schechter \citep{1976ApJ...203..297S} function. The thin lines are
    similar models, but with truncations at $10^6\,\msun$ (higher curve)
    and $10^4\,\msun$ (lower curve).}

    \label{fig:lf} 
\end{center}

\end{figure*}

Following the source selection as described in \S~\ref{sec:data}, we
generate the LF of all clusters in the three available filters (left
part of Fig.~\ref{fig:lf}). The three columns on the left correspond
with the \B,\,\V\, and \I\, filters. In the top panels we fit a double
power-law distribution, with a variable location of the break and
indices and in the bottom panels we fit a Schechter function with
variable $M_*$ and index. The results of the different fits are
summarized in Table~\ref{tab:obs}. Note that the slopes on the faint
sides correspond to the typical value of $-2$, found in other
galaxies. The slopes on the bright side ($\alpha_2$) are much
steeper. The break appears at higher luminosities in the \I\
filter. The double power-law break shifts about 0.5 mag to the red,
going from \V\, to
\I. The $M_*$ values from the Schechter fit are about 0.2 mag brighter
in
\I\, as compared to \B\, and \V.

 In \citet{gieles05b} we constructed the LF of M51 clusters based
on the data of \citet{2005A&A...431..905B}. The data was based on two
{\it WFPC2} pointings, and the LF showed that a double power-law
function was better fit than a single power-law by a factor of two,
when comparing $\redchisq$ values. In this case, we obtain $\redchisq
= 0.78$ for the double power-law fit in the \V\ band, whereas the
single power-law fit results in $\redchisq = 203$. This shows that the increase
of the field-of-view and depth of the current data set contributes
considerably to the significance of the result.

The downturn at bright magnitudes and the slopes of the double
power-law distribution are very similar to what
\citet{1999AJ....118.1551W} found for the LF of the ``Antennae''
galaxies. The bend and $M_*$, however, occur about 1.6 mag brighter
in the ``Antennae'' LF. We will explain this behavior of the LF and
the difference between the ``Antennae'' and M51 in \S~\ref{sec:model}.

\begin{table}[!t]
\caption{Results of the different function fits to the observed LFs of Fig.~\ref{fig:lf} and the simulated LFs of \S~\ref{sec:model} .}
\vskip -1mm
\begin{center}
\begin{tabular}{llllll}
\hline\hline
        & \multicolumn{3}{l}{Double power-law} &
\multicolumn{2}{l}{Schechter} \\ Filter \hspace{0.5cm} &$\alpha_1$
&$\alpha_2$ &Break\hspace{0.5cm} &$\alpha$ &$M_*$ \\\hline

  \multicolumn{6}{c}{Observations} \\\hline
F435W &   1.94  &    2.72  &   -8.88  &    1.81  &   -9.87  \\
F555W &   1.93  &    2.75  &   -8.93  &    1.80  &   -9.85  \\
F814W &   1.93  &    3.08  &   -9.43  &    1.78  &  -10.05  \\

  \multicolumn{6}{c}{Models with $M_C = 10^5\,\msun$} \\\hline
F439W &   2.01  &    2.76  &   -8.99  &    1.85  &   -9.89  \\
F555W &   1.93  &    2.71  &   -8.94  &    1.79  &   -9.94  \\
F814W &   1.90  &    2.70  &   -9.60  &    1.76  &  -10.57  \\
\hline\hline
\end{tabular}
\end{center}
\label{tab:obs}
\vspace{-0.5cm}
\end{table}

\section{Comparison with a cluster population model}
\label{sec:model}

We have developed an analytical model which can reproduce the
observable properties of a cluster population. The model was
introduced in \citet{2005astro.ph..6066G} for comparison with the age
and mass distribution of M51 and used in \citet{gieles05b} to compare
with LFs in different galaxies. The model generates  cells which
are equally spaced in log(age/yr) and log(mass/\msun). Weights are
assigned to each cell, such that the integrated weights as a
function of mass correspond to a chosen CIMF and formation rate and in
that way correspond to number of clusters. Working with weigths is
preferred over generating the actual number of clusters, since this
can be a very high number.  Then cells are evolved  as if they
were clusters, where mass loss due to stellar evolution and
disruption are taken into account by analytical functions derived by
\citet{2005astro.ph..5558L}.

In \citet{gieles05b} we showed that a physical limit to the
maximum cluster mass will cause a bend in the LF. For details we refer
to that paper. We generated several populations with different maximum
masses. A constant cluster formation rate between $6\times10^6\,$yr
and 10 Gyr is assumed and a power-law CIMF with  an exponential
cut-off $N\,\dr M \propto M^{-\beta}\,\exp(-M/M_C)\,\dr M$, with $\beta=2$,
similar to what has been proposed for the CIMF of globular clusters
\citep{2000ApJ...542L..95B}.  The disruption parameters from
\citet{2005astro.ph..6066G} are adopted (i.e. $t_{\rm dis} = 10^8\,
(M_i/10^4\,\msun)^{0.62}\,$yr). We note that with these disruption
parameters all cluster older than 2 Gyr are removed from the sample.
We applied an extinction of $A_V = 0.25\,$mag to all clusters, which
is close to the average extinction measured by
\citet{2005A&A...431..905B}. All masses are converted to {\it
ACS} filter magnitudes, depending on their age, using the {\it
GALEV} simple stellar population models
(\citealt{2003A&A...401.1063A}; \citealt{2002A&A...392....1S}). We fit
the same functions to the artificial LFs as we did to the data. In the
right panel of Fig.~\ref{fig:lf} we show the result for the \V\,
filter for $M_C = 10^5\,\msun$. The predicted LF parameters for all
filters are summarized in the bottom part of Table~\ref{tab:obs}. The
bright side shows a steeper slope than the faint part, as is the case
in the data. This is because the clusters with the maximum mass fade
and therefore have an age-dependent contribution to the LF. The
brightest cluster will be the youngest and going faint-wards, more
older clusters will contribute to the LF.  Faint-ward of $M_V
\simeq -9$, clusters of all ages contribute. The fact that we find
shallower slopes on the faint side than the input MF slope is due to
the mass dependent disruption. 
The fact that the bend occurs at a higher luminosity in the
\I\ filter, in both the model and the data, is an important
confirmation of our idea that the bend is caused by a truncated MF.

When we make a direct comparison with the ``Antennae'' galaxies, the
MF is truncated at a higher mass ($M_C = 4\times10^5\,\msun$), since the
bend there occurs 1.6 mag brighter \citep{1999AJ....118.1551W}.


\section{Discussion and conclusions}
\label{sec:conclusions}


Does the fact that other galaxies have a single power-law LF imply
that there is no upper limit to the cluster mass there? Probably
not. The CIMF has to be sampled well enough to reach the critical
\mmax, only then a bend will show up  in the total cluster LF. M51 and the
``Antennae'' are forming enough clusters such that the CIMF is sampled until
\mmax, and a bend in the LF is observable. The difference between
\mmax\ in M51 and the ``Antennae'' galaxies suggest that \mmax\ is
environment dependent. 

These environmental differences might be caused by variations
in the giant molecular cloud (GMC) mass
distribution. \citet{2003ApJ...599.1049W} shows that the cloud mass
function of the ``Antennae'' galaxies is truncated at higher masses
than that of M51, which in turn is at much higher masses than in the Milky Way
\citep{1997ApJ...476..166W}. Other galaxies also seem to have GMC mass
distributions which are truncated at the high mass end
\citep{2005astro.ph..8679R}.  A truncated GMC mass
function might impose a physical limit to the maximum star cluster
mass, which will be observable in galaxies with a high star/cluster
formation rates.

\begin{acknowledgements}
We thank Marcelo Mora at ESO/Garching for kindly providing us with the
empirical ACS PSFs.
        
\end{acknowledgements}

\bibliographystyle{aa}

\end{document}